# Providing a Periodic Control Solution for Balance Control While Standing Using a Pendulum-Based Approach

[1]Golnoush Shahraki*, [2]Majid Sadrzadeh, [3]Elyas Irankhah

1- Department of Biomedical Engineering, Islamic Azad University of Mashhad, IR
2- Assistant Professor of Emergency Medicine, Department of Emergency Medicine, Mashhad University of Medical Sciences, Mashhad, IR
3- Department of Mechanical Engineering, University of Massachusetts Lowell, MA

*Corresponding Author: Shahraki.golnush@gmail.com

## Abstract

The stability of standing in humans is a complex process that leads to maintaining the upright position against external disturbances. Balance control during standing is of vital importance for humans in daily life. An issue that is still not clearly understood is which control mechanism the central nervous system uses to maintain stability. In the rehabilitation of standing function, the coordination pattern between the angles of the leg joint of a healthy person should be restored. For example, one of the rehabilitation methods is functional electrical stimulation. In the work that was mainly done in the control of standing balance with functional electrical stimulation, the problem of the optimal pattern using the phase space was not mentioned at all, and a series of predetermined desired curves were assigned to the joints, and the controller only used these curves. followed, while the origin of these curves are not real patterns. Therefore, the main goal of this project is to design a periodic controller based on phase space. In such a way that a mapping related to standing is detected first, then a feedback controller is designed so that it is activated only when the system state space curves find a significant distance from the detected mapping, then the feedback controller is activated, and it adjusts the control signal so that the system state space curves come close to the detected mapping.

## Introduction

In the realm of balance control and human biomechanics, several pivotal studies have laid the groundwork for understanding and improving balance mechanisms. In [1] researchers propose an innovative model and control approach. This study transcends the limitations of traditional Linear Inverted Pendulum Model (LIPM) methods by incorporating ankle joint variables into the control of humanoid robots' standing balance, an approach that could significantly influence robotic technology in rehabilitation and assistive devices. Simultaneously, the study [2] by Cabrera JL and Milton JG explores into the inverted pendulum model, a cornerstone in discussing human balance during standing and locomotion. This research examines three experimental paradigms of time-delayed balance control: a mechanical inverted time-delayed pendulum, stick balancing at the fingertip, and human postural sway during quiet standing. Their findings underscore the complexity of balancing as a time-dependent state, challenging the conventional notion of a simple fixed-point attractor.

Further contributing to this field, the paper [3] explores the role of predictive feedback mechanisms in pendulum balancing, an aspect significant to understanding human balance control. This is complemented by [4] which sheds light on the use of control systems in maintaining balance during standing, a vital function in human mobility.

Expanding this interdisciplinary approach into the realm of medical diagnostics, the papers [5] venture into the application of AI methodologies in neurological disorder diagnosis. Similarly, [6] evaluates specific diagnostic tools like MRI and EEG, enhancing the precision of epilepsy diagnosis. These studies underscore a critical convergence of biomechanics, robotics, and medical diagnostics, highlighting the potential of advanced technologies in both understanding human balance and improving health outcomes in neurological conditions.

Another significant contribution is made in [7], where researchers explore the integration of virtual reality (VR) in balance and gait training. This paper highlights how VR can simulate various environments and conditions, offering a safe and controlled setting for patients to improve their balance and mobility skills.

Furthering the discussion on neurological disorders, [8] presents an in-depth analysis of brain network connectivity in epilepsy patients. This study uses advanced neuroimaging techniques to unravel the complexities of brain function alterations in epilepsy, paving the way for more targeted treatment approaches.

By designing a periodic controller based on phase space, as outlined in the project's main goal, there is potential to revolutionize the way balance is understood and maintained. This controller aims to first detect a mapping related to standing, and then activate a feedback mechanism only when significant deviations from this mapping are observed. Such an approach not only aligns with the findings from the aforementioned studies but also promises to bring us closer to restoring natural balance patterns in individuals, enhancing their quality of life and mobility.

## 1. Method & Materials

The main objective of this research is to provide a phase-plane COP-based control strategy for describing balance preservation. Considering COP information regarding movement, this study will analyze the phase-space COP information. The proposed control method will utilize a behavior-based controller driven by phase-plane characteristics. This involves identifying the mapping associated with standing by plotting the phase space, followed by designing a feedback controller that only becomes active when the phase-space curves significantly deviate from the identified mapping. The feedback controller is then activated, adjusting the control signal to bring the system's phase-space curves close to the identified mapping. The research method will be elaborated in detail below.

## 1-2. Dataset

Due to the unavailability of laboratory conditions, data was collected from the PhysioNet database[1]. The dataset includes recordings of force platforms from individuals undergoing stability tests. The recorded data are from individuals who were completely healthy and had no skeletal, motor, or cognitive impairments. Subjects performed the standing posture under four different conditions: with eyes open or closed, while standing on a firm or unstable surface. Each condition was tested three times for each individual in the experiment. In total, 1930 trials performed by 163 different subjects are included in this database. Each recording lasted for one minute, with a sampling frequency of 100 Hz, and pre-processing was done using a 0.1 Hz low-pass filter. Signal files consist of 8 channels with 3 different recording types as follows:

1. N Force (X, Y, Z) , 2. Time nm (X, Y, Z) , 3. Center Pressure cm (X, Y , Z)

The signal file name is in the format XXXX0BDS, where XXXX is the experiment number. The following qualitative tests were used for each subject:

- ✓ Short Falls Efficacy Scale International Version
- ✓ Short International Physical Activity Questionnaire
- ✓ Trail Making Test
- ✓ Small Balance Assessment System Tests

Additionally, participants were interviewed about certain socio-cultural, demographic, and health-related information, including age, medications, and medical conditions. The collected information and test results are presented in BDSinfo.txt and BDSinfo.xlsx files. Each record's header contains some of this subject-specific field information.

## 1-3. Phase Space Formation

In the theory of dynamic systems, a phase space is a space in which all possible states of a system are represented, with each possible state corresponding to a unique point in the phase space. For mechanical systems, the phase space typically comprises all possible values of the position and velocity variables.

As per our study, the primary objective is to propose a controller based on phase space information of the Center of Pressure (COP) for describing balance preservation. Therefore, as a first step, we will analyze the COP phase space information. To obtain the phase space corresponding to each subject, $COP_x$ and $COP_y$ data are plotted against each other. The resulting curves in the phase space are termed phase space trajectories. The phase space trajectory related to the collected data is illustrated in the following figures:

---

[1] https://physionet.org/

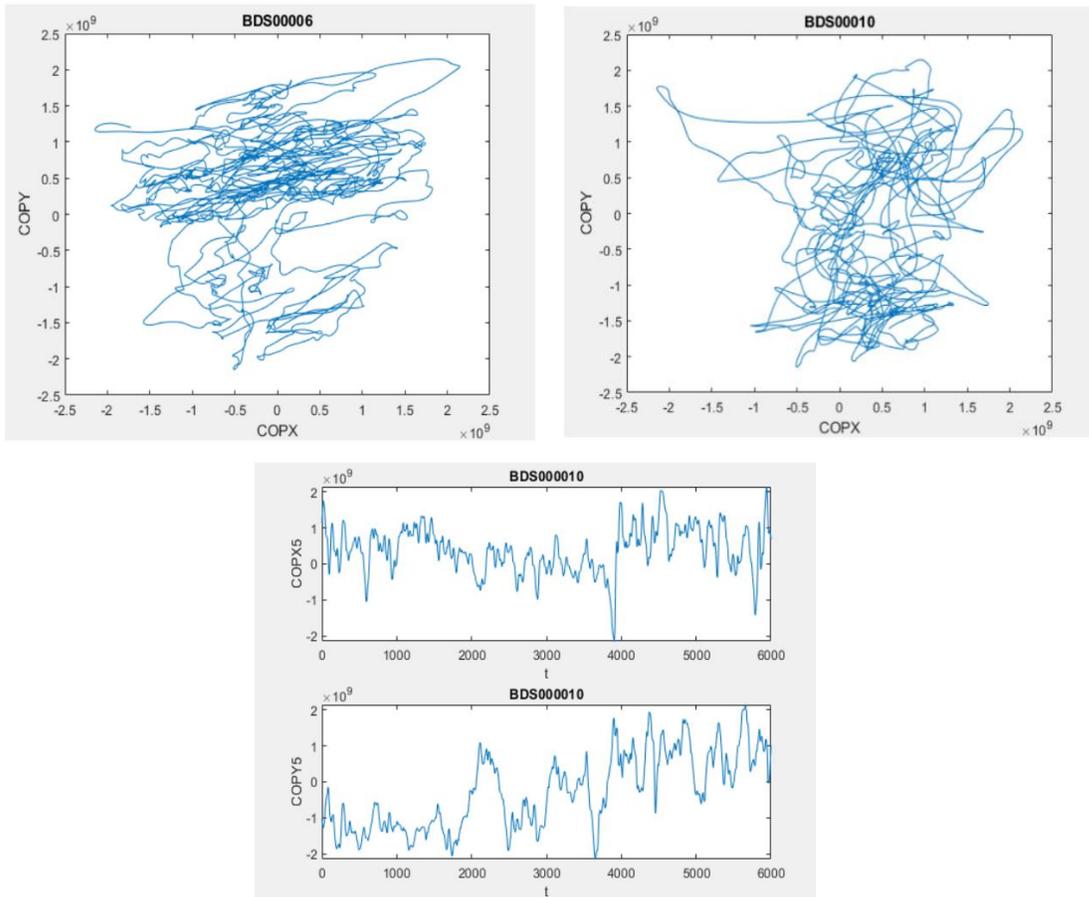

**Fig1.** Phase space trajectories in BDS00006 and BDS000010

## 1-4. Identification of Poincaré Sections and Mapping

Poincaré sections are a solution for selecting points in phase space that contain significant information about the dynamics of a system. The use of dynamic-based methods, such as Poincaré sections, which utilize the coordinates of intersections in phase space, can be instrumental in detecting the dynamics of biological systems. It provides a pathway to gain a better understanding of the complexity of nonlinear dynamic systems. In this phase, the objective is to find the equation of a curve based on the attractor set.

In this research, points are chosen in the following manner:

- Where $COP_x$ is at its maximum
- Where $COP_y$ is at its maximum
- Where $COP_x$ is at its minimum
- Where $COP_y$ is at its minimum
- The point (0, 0)

Therefore, by selecting these five points, we gather a set of data points that need an appropriate equation and mapping to describe the relative changes among them. In fact, we must identify the model parameters that can effectively capture the variations observed among these points.

## 1-5. Logistic Mapping

One of the most popular chaotic functions is the logistic map, also known as the Verhulst model. The logistic equation, which models population growth, is represented as follows:

$$X_{n+1} = rX_n(1 - X_n) \quad (1)$$

Where, $X_{n+1}$ is the value at the next time step, $X_n$ is the current value and $r$ is the control parameter that determines the behavior of the system.

The logistic mapping, also known as the logistic function, is a polynomial mapping that is often used as an example to illustrate how complex chaotic behavior can emerge from simple nonlinear dynamical equations. The logistic map is an irreversible mapping, meaning it can be iterated forward in time from $X_n$ to $X_{n+1}$, but reversing this process is not generally possible. This mapping is also referred to as a recurring mapping because it maps one value of X (e.g., $X_0$) to another value (e.g., $X_1$), and through continuous iterations, it exhibits a wide range of behaviors. The sequence of iterated values forms a path or trajectory, often called an orbit.

## 1-6. Controller Implementation

The control mechanism employed in this research is periodic, meaning that the designed controller only becomes active when the distance between the phase space curves of the system and the identified second-degree curve exceeds a certain threshold. In this phase, the minimum Euclidean distance between the phase space curve and the fitted second-degree curve is measured for each point. If this distance exceeds the threshold, the selected fuzzy feedback controller becomes active, or alternatively, the fuzzy system activates the PID controller, which generates a control signal that modulates the intensity of muscle stimulation involved in maintaining the standing posture. The PID controller is configured to adjust the system's state space curves to closely match the identified mapping [9]

If the distance between the system's state space curve and the identified mapping is less than the threshold, the fuzzy controller remains inactive, and no muscle stimulation is applied. In this study, the threshold value for activating the controller has been determined through trial and error. The designed control system includes both a fuzzy controller and a PID controller.

## 1-7. PID Controller Structure

PID controllers are designed based on microprocessor technology, and different industries may use various PID structures and equations. The PID equations commonly used include the parallel, ideal, and series forms.

In the parallel PID equation, the proportional, integral, and derivative actions operate separately and have a combined effect on the system. In the ideal PID equation, the gain $K_p$ is a constant that is distributed evenly across all the terms in the equation. Therefore, changes in $K_p$ will have an impact on the other terms of the equation.

In the series PID equation, the proportional gain $K_p$ is distributed across all terms of the ideal PID equation, but in this equation, the integral and derivative constants affect the proportional performance. The block diagram for this series PID controller is depicted in the following figure:

In this research, we used a parallel PID controller. To achieve the desired output, this controller needs to be properly tuned. The process of obtaining an ideal response from the PID controller by adjusting the PID parameters is called PID tuning. PID tuning involves adjusting the values of the proportional gain $K_p$, derivative gain $K_d$, and integral gain $K_i$ optimally. There are various methods for tuning PID controllers to obtain the desired response, which will be explained below:

- Trial and Error Method
- Process Reaction Curve Technique
- Ziegler-Nichols Method
- Relay Method
- Using Software

Technically, we applied the Ziegler-Nichols method to tune the parameters $K_p$, $K_i$, and $K_d$. The obtained values are as follows:

- $K_d$ = 0.93
- $K_i$ = 1
- $K_p$ = 0.87

## 1-8. Fuzzy System Design

The fuzzy controller designed in this research has one input and one output. The input of the fuzzy system is a parameter called "d," which represents the distance between the phase space trajectories of the system and the identified second-degree curve at each moment. The fuzzy system employs very simple fuzzy rules. For example, one of the rules could be: "If the system is active (y), then the value of 'd' is big."

In our controller, we initially receive the parameter 'd,' which is essentially the Euclidean distance of the trajectory to one of the points on the mapping. If the size of the distance between the phase space trajectory of the system and the identified mapping corresponding to the signal is less than a threshold, the fuzzy controller will be inactive. Otherwise, it will become active. As mentioned earlier, the threshold value for activating the controller is determined through trial and error, resulting in a value of 0.05 in this research.

To design the fuzzy system, we utilize a Mamdani-type inference system with one input ('d') and one output ('y'). The designed fuzzy system, the membership functions for the input 'd,' and the fuzzy rules for the system are depicted in the following figures.

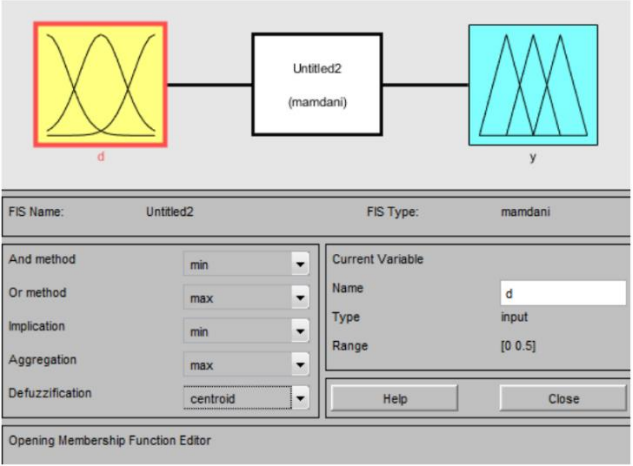

**Fig2.** The designed fuzzy system

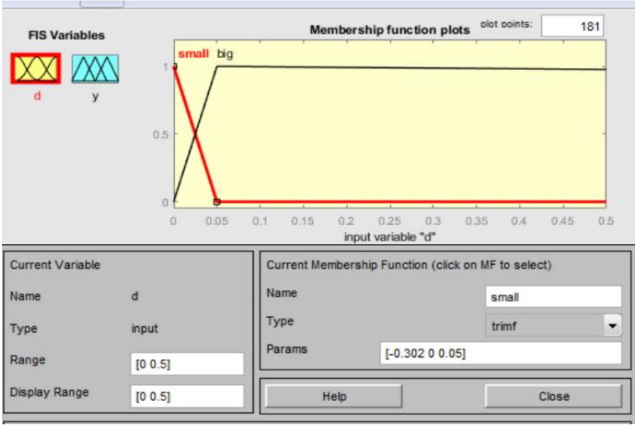

**Fig3.** Input membership functions d for fuzzy controller

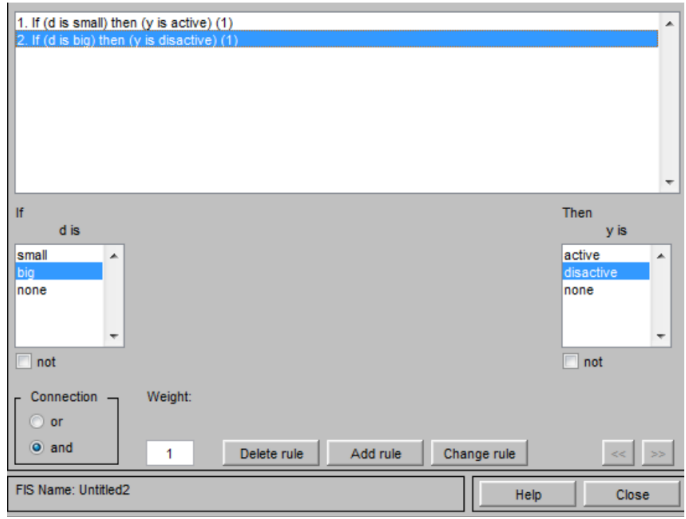

**Fig4.** Fuzzy rules of the designed fuzzy controller

## 2. Results

Some studies recently introduced specially [10] a new stability criterion that could be used to assess an individual's standing posture based on center of pressure (CoP) measurements. They demonstrated that during quiet standing, the CoP moves between the feet both in the anterior-posterior and medial-lateral directions and can be found in a very small elliptical region, which they referred to as the "high preference zone," in 99% of cases. As long as the CoP remains within this zone, the best state of balance is achieved. If the CoP moves out of the high preference zone but remains within the "preference low" zone, acceptable balance is still maintained. However, once it exits the preference low zone, balance is compromised.

Without statistical error, it can be assumed that all three boundaries of the stability zones have a common center on the midline, which starts from the heel and extends towards the toes, encompassing 47% of the foot's length. The major and minor diameters of the elliptical zones are functions of the individual's foot sole and can be obtained from the following equations.

(2)

- High Performance Zone: $\frac{(x-0.47)^2}{d1_{hp}^2} + \frac{y^2}{d2_{hp}^2} < 1$
- Low Performance Zone: $\frac{(x-0.47)^2}{d1_{hp}^2} + \frac{y^2}{d2_{hp}^2} \geq 1$ and $\frac{(x-0.47)^2}{d2_{hp}^2} + \frac{y^2}{d1_{hp}^2} < 1$
- Undesirable Zone: $\frac{(x-0.47)^2}{d2_{hp}^2} + \frac{y^2}{d1_{hp}^2} \geq 1$ and $\frac{(x-0.47)^2}{d2_{hp}^2} + \frac{y^2}{d1_{hp}^2} < 1$
- Unstable Zone: $\frac{(x-0.47)^2}{d2_u^2} + \frac{y^2}{d1_u^2} \geq 1$

The values of 1st and 2nd are shown in the table below, and these values have been normalized. In this study, we considered the length of the individual's foot sole as 20 cm, and by multiplying this number by the values in the table, we obtain hp1d, hp2d, and other parameters in the equations above.

**Table1.** The acquired variables

| Zone boundaries | d1 | d2 |
|---|---|---|
| High Preference | 0.16 | 0.07 |
| Low Preference | 0.57 | 0.43 |
| Undesirables | 0.97 | 0.59 |

If the disturbance introduced to the individual is at a level where the ankle can still maintain control, and the COP moves from the preference high to preference low zone, the balance condition is still acceptable, and the controller remains inactive. However, if the disturbance is significant enough that the ankle alone cannot maintain balance, the individual enters the "Undesirables" zone, and the controller becomes active. Therefore, based on changes in $COP_y$ and $COP_x$, the status of the control signal should be examined to determine the degree of synchronization between them.

## 3. Conclusion

The main objective of this research is to present a phase-based controller using COP (Center of Pressure) information to describe balance preservation. The designed fuzzy controller only becomes active when the distance between the phase space curves of the system and the identified signal mapping exceeds a certain threshold. With the activation of the fuzzy system, the PID controller generates a control signal, which is the intensity of muscle stimulation involved in maintaining the standing performance. It adjusts this signal in a way that brings the state-space curves of the system closer to the identified mapping.

If the distance between the phase space curves of the system and the identified signal mapping is less than the threshold, the fuzzy controller remains inactive, and no stimulation is applied to the muscles. However, if the disturbance is significant enough that the ankle alone cannot maintain balance, the individual enters the "Undesirables" zone, and the fuzzy controller becomes active.

To evaluate the controller, a new stability metric is used, which assesses an individual's standing status based on measurements of the Center of Pressure (CoP). During calm standing, the CoP moves between the back-and-forth and left-and-right positions between the two feet. In this approach, three zones are defined to examine an individual's balance during standing: "preference high," "preference low," and "Undesirables." As long as the CoP remains within the "preference high" zone, the best balance condition is maintained. If the introduced disturbance is manageable, and the CoP moves from the "preference high" zone to the "preference low" zone, the balance status remains acceptable, and the controller remains inactive. However, if the disturbance reaches a level where the ankle alone cannot maintain balance, the individual enters the "Undesirables" zone, and the fuzzy controller becomes active.

Therefore, the status of the control signal should be examined based on changes in $COP_y$ and $COP_x$. In other words, changes in COP should match the controller's output, which is the intensity of muscle stimulation, and whenever the COP deviates from the balance boundary, the control signal should increase. The graphs below show the $COP_x$ and $COP_y$ plots for an individual, as well as the output control signal of the designed controller.

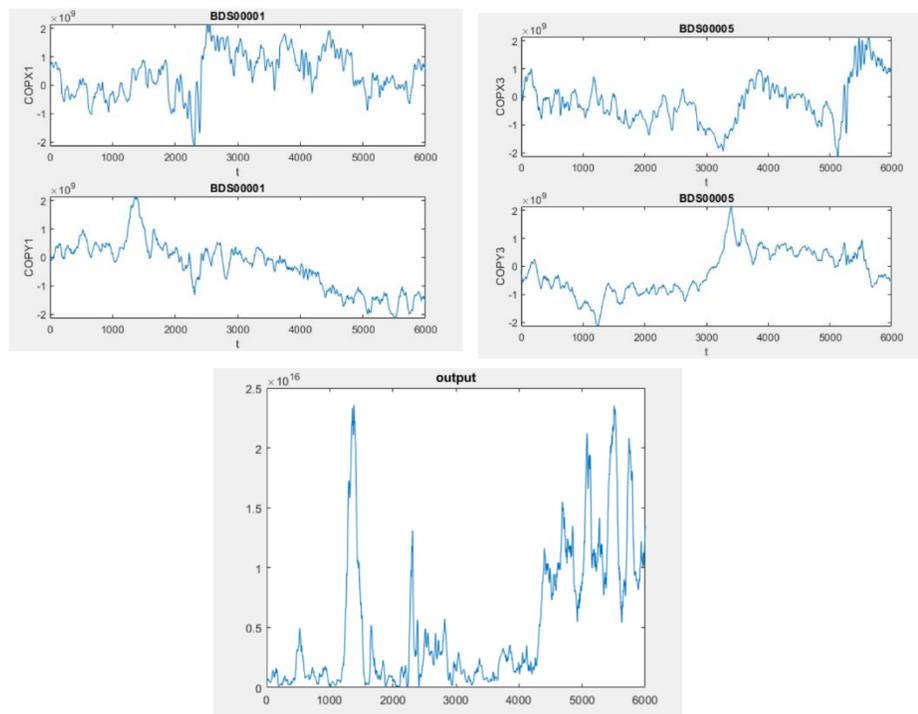

**Fig5.** $COP_y$ and $COP_x$ variation curve over time for data BDS00001 & the output of controller signal